\documentstyle[epsfig]{aipproc2}
\begin{document}
\title{T=0 Neutron-Proton Correlations at high angular momenta\footnote{Invited talk at the
International Conference on Nuclear Structure, August 10-15, Gatlinburg, USA}}

\author{R. Wyss}
\address{Royal Institute of Technology (KTH), Physics Department Frescati,
Frescativ.~24,
S-104 05 Stockholm}

\maketitle

\begin{abstract}
The properties of $T\!=\!0$ neutron-proton 
correlations are discussed within the
framework of different model calculations. Single-j shell calculations
reveal that the 
$T\!=\!0$ correlations remain up to the highest frequencies. They are 
more complex than the $T\!=\!1$ corrlations 
and cannot be restricted to $L\!=\!0$
pairs only. Whereas it may be difficult to find clear evidence for
$T\!=\!0$ pairing at low spins, 
$T\!=\!0$ correlations
are found to induce a new excitation scheme
at high angular momenta.

\end{abstract}

\section*{Introduction}

Pairing correlations have always played a decisive role for 
the low energy structure of atomic nuclei. 
In close analogy to the BCS-theory of superconductivity, 
it was suggested
early on that the ground state of most nuclei is formed
by a coherent superposition of pairs of nucleons, moving in time
reversed orbits.\cite{Boh58} 
Although the BCS-theory of the nuclear 
pairing interaction is charge independent,
one in general considers only scattering between pairs of
like particles (neutron-neutron and proton-proton).

Atomic nuclei are composed of two different kind of fermions
which can occupy identical states. This is an unique situation, 
that is not found in other
fermionic systems. Since the short range nuclear force 
is attractive, 
nuclei in close vicinity of the $N\!=\!Z$ line may form a condensate
of neutron-proton (np) pairs that have totally symmetric wave-functions
in spin and ordinary space.  
Different approaches to generalize the pairing interaction 
to fully take into account the iso-spin degree of
freedom have been discussed in the literature in the
60ties and 70ties, see e.g.
Ref.~\cite{Goo79} and references therein. 
Following
the experimental observation of $^{100}$Sn and the spectroscopic
study of a large number of heavy $N\!=\!Z$ nuclei in recent years,
the quest for neutron-proton pairing has gained
renewed interest.\cite{Eng96,Sat97,Goo99}
Indeed, the massdefect at the $N\!=\!Z$-line, the so called Wigner
energy finds a microscopic explanation in a generalized pairing
theory\cite{Sat97}.
In the following we use the standard notations
for isospin ($t,~T$), intrinsic spin ($s,~S$), orbital
angular momentum ($l,~L$) and total angular momentum
($j,~J$), where small letters denote the single
particle content and capital letters the vector added quantities.


Pairs of particles moving in time-reversed orbits
have the largest momentum exchange when their
total spin $S$ and angular momentum 
$L$ are coupled to zero. 
Such pairs form a triplet in iso-space $T\!=\!1$, 
being decomposed of either proton (neutron-) pairs with
$T_z\!=\!\pm1$ or a neutron-proton  pair with
$T_z\!=\!0$. Strong short range correlations of particle-particle (pp)
type, are in general treated by means of
the standard 'monopole' pairing force ($T\!=\!1$) 
with $T_z\!=\!\pm 1$ pairs only.
In order to treat protons and neutrons on the same footing,
one may extend the formalism to include $T_z\!=\!0$ pairs.\cite{Gos65}
For even-even nuclei, such extensions 
may be considered less interesting, 
since the $T_z\!=\!0$ pairs are simply related to 
$T_z\!=\!\pm 1$ pairs via rotation
in iso-space.
The binding energy of even-even nuclei e.g. is not
affected by the inclusion of $T_z\!=\!0$ pairing\cite{Sat97}.
However, for odd-odd $N\!=\!Z$ nuclei, many interesting properties
may emerge, that will be addressed in a forthcoming paper\cite{Sat99}.

\begin{figure} 
\vspace*{13.0 cm}
\hspace*{2.5 cm}
\includegraphics{fig: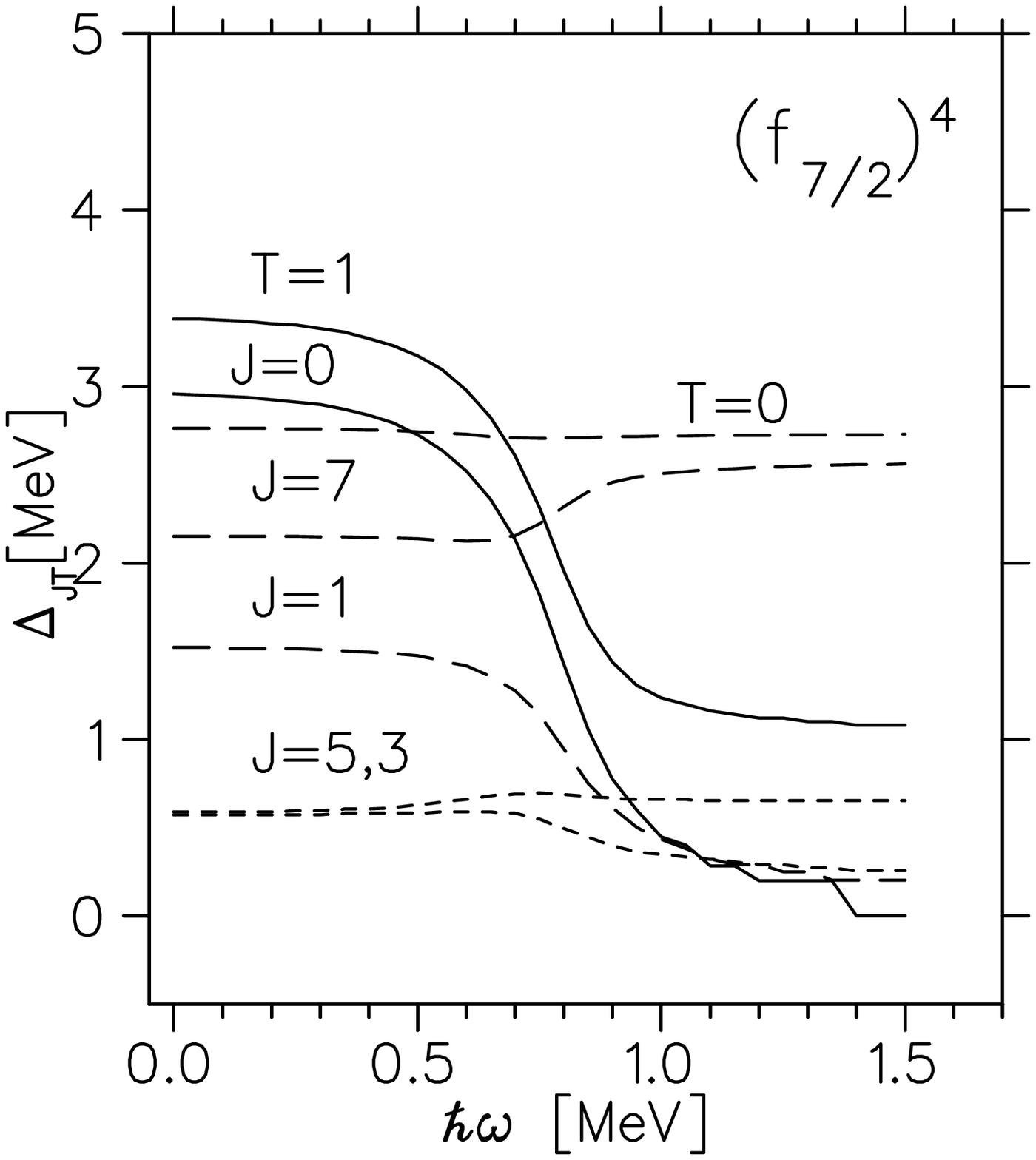}
\noindent
\hspace*{2.cm}
\vspace*{-4.cm}
\caption{Decomposed ($J,~T$) pairing energy for 2 protons and 2 neutrons in the
$f_{7/2}$ shell. The results for 4 protons and 4 neutrons is
similar. }
\vspace*{10pt}
\label{fig1}
\end{figure}

In nuclear physics it is well known that
the Deuteron is bound only in the spin triplet, $S\!=\!1$
iso-spin singlet, $T\!=\!0$ state. 
Also nucleon-nucleon scattering data show 
stronger attraction for the spin triplet than
spin singlet state.
One may therefore expect that neutron-proton pairing is more
important at the $N\!=\!Z$ line than neutron-neutron and 
proton-proton pairing.

In the context of the shell model, one defines a 'monopole' 
pairing interaction for the $L\!=\!0$, $T\!=\!1$ singlet state\cite{Tal62}, 
and one may proceed in an analoguous manner 
for the $T\!=\!0$, $L\!=\!0$ and $S\!=\!1$ triplet state.
This restricted definition is of advantage for
algebraic models \cite{Dus86,Eva81,Isa97} since
it limits the model-space
when analyzing properties of np-correlations.
It may be appealing, to
consider $L\!=\!0$ pairs only for
the pairing interaction.
However, this neglects several 
important aspects. 
Whereas the $L\!=\!0$ part of a short range $T\!=\!1$ interaction
like the $\delta$-force is dominant, it plays 
only a minor role for the $T\!=\!0$ interaction\cite{Tal62}.
This is nicely seen in the experimental
spectra of odd-odd nuclei, where the low-lying $0^+$-state
is in close vicinity 
to states with odd spins and even parity
corresponding to the $(j^2)_{J\!=\!1}$ and $(j^2)_{J\!=\!2j}$ configuration.

For mean field calculations, 
it does not make much sense 
to restrict
to a $L\!=\!0$ pairing force, especially not
for the $T\!=\!0$ channel. 
Already the 
$T\!=\!0$ pair with lowest $J,~J\!=\!1$ has a
strong $L\!=\!2$ component that needs to be taken into account.
Not even the $T\!=\!1$ pairing channel can be restricted
to $L\!=\!0$-pairing, when one aims at a quantitative
description of  the rotational motion and pair-breaking mechanism in
deformed nuclei. One has to include at least 
$L\!=\!2$ ('quadrupole'-)pairing\cite{Die84,Sat94}. 
The contribution to the binding energy steming from the $L\!=\!2$-pairing
may be negligible 
small, of the order of a tens of $keV$, but 
it is certainly coherent\cite{Sat94}.

In the present paper, we investigate the influence of
the $T\!=\!0$ correlations on properties of rotational
bands at high angular momenta. Single-j shell calculations
are presented in section 1 and mean-field calculations
based on the HFB-method in section 2.


\section*{Neutron-proton correlations in a single-j shell}

In a single-j shell, one can diagonalize exactly the two-body
interaction and also investigate the dependence 
of different correlations on angular
frequency\cite{Shei90}. The pair-gap, is not defined
in such a model.
Nevertheless, 
one may 
decompose the expectation value of the
two-body interaction, $E_{exp}$, 
with respect to the contributions coming from the different angular
momenta $J$.
We have performed a single-j shell calculation for different number of
particles in the $f_{7/2}$ shell
and relate a 'gap'-parameter $\Delta_{JT}$ to the expectation
energy via the following equation
\begin{equation}
E_{exp} = { 1 \over 2} \sum_{JT} { \Delta_{JT} \Delta_{JT}^{*} 
\over E_{JT}},
\end{equation}
where $E_{JT}$ is the value  of
the normalized, anti-symmetric two-body matrix elements of the
$\delta$-function interaction\cite{She99}.
As expected, the single-j shell calculations indeed result in
a strong binding for 
maximum and minimum $J$ in the $T\!=\!0$  channel, but also 
shows that states
with intermediate $J$ are important, see fig.~\ref{fig1}.
Whereas the binding for the $T\!=\!0$-pairs
is rather complex, the $T\!=\!1$ part of the interaction
is dominated by a single $J$.
It is the dominant role of the $J\!=\!0$ pairing in the
$T\!=\!1$ channel that may justify the
restriction to $L\!=\!0$ in that channel. These
properties of the $T\!=\!1$ and $T\!=\!0$ correlations have
been known since
long\cite{Tal62,Sch76}, see also the recent investigation 
in\cite{Sat97a}.
When we start to rotate the nucleus,
the Coriolis force tends to align the quasi-particles along the
rotational axis, resulting in
the well known pair breaking mechanism. 
The $T\!=\!1$ pairing becomes reduced
and the $L\!=\!0$ part totally
quenched. Since the $J\!=\!1$ part of the $T\!=\!0$ correlations
also involve particles in which the intrinsic $j$'s are
mainly coupled antiparallel, this part is reduced in
a similar fashion as the $J\!=\!0$ correlations. At the same time,
the $J\!=\!7$ correlation pick up in strength and compensate the
loss from the $J\!=\!1$. 
Although there exist only one $J\!=\!7$ state
in the $f_{7/2}$ shell, this state has the highest
degeneracy, $2J+1$ , implying that many pairs may contribute 
coherently to this state. 
The total $T\!=\!0$ correlations remain 
unchanged as a function of frequency. Apparently, the $T\!=\!0$ 
energy is
not affected by rotation and indeed, we do not observe 
any effect on the alignment by switching on or off that
part of the interaction.

Recent shell model calculations 
question the importance of $T\!=\!0$ correlations 
for the spectrum of $^{48}$Cr\cite{Pov98}.
However, the only matrix elements considered in this
analysis are those of $L\!=\!0$.
As shown above, in contrast to the $T\!=\!1$ channel, 
the  $T\!=\!0$ interaction
is most attractive for parallel $J\!=\!2j$ and
antiparallel $J\!=\!1$ coupling of the individual angular momenta $j$,
Fig.~\ref{fig1}. Even the
$T\!=\!0,~J\!=\!1$ matrix element
contains contribution from both $L\!=\!2$ and $L\!=\!0$.
The requirement of total $L\!=\!0$ and
$S\!=\!1$ for a $l^2$-pair of $T\!=\!0$
implies that the individual projection of the orbital
angular momentum, $l$,
($l_m$) have to be antiparallel and
the individual spins parallel.
For a given $l^2$ configuration, one may couple
the individual orbital angular momenta and intrinsic spins $s$, to a total
angular momentum $j,~j\!=\!l\pm 1/2$. When decomposing the contribution
to the $L\!=\!0$ multiplet 
in terms the 'spin-orbit' partner pairs 
$(j\!=\!l+\frac{1}{2},~j\!=\!l-\frac{1}{2})$,
pairs with $(j\!=\!l+1/2)^2$ and
$(j\!=\!l-1/2)^2$, respectively,
we find from the appropriate $6j$ symbols
a ratio of 8:7:2 (for $l=2$).
This implies of course, that the
$L\!=\!0$, $T\!=\!0$ pairing becomes strongly reduced
by the nuclear mean-field in the presence of the
spin-orbit interaction. 
The concept of a restricted $L\!=\!0$, $T\!=\!0$ pairing may
be meaningful up to the sd-shell, where the $LS$-coupling
still has validity. When entering the
$f_{7/2}$ shell, we do not expect the
$L\!=\!0$-pairing in the $T\!=\!0$ channel to play a very
important role. Therefore, it is not at all surprising
that this interaction turns out to have little influence on
the spectrum of $^{48}$Cr.

One should remember, that the definition of the
pairing gap originates from the mean-field approximation
where it defines the average gap or potential 
felt by a pair of particles at the Fermi-surface\cite{Boh58,Bog58,Bel59}.
The pairing interaction is thus intimitely linked to the symmetry-breaking
of the mean-field approach,
which strictly separates between a particle-particle (pp) and 
particle-hole (ph) field. In contrast,
the two-body 'pairing' matrix element of the shell-model is a different entity
containing both particle-hole and particle-particle part
and one has to be carefule when comparing results of
the two different approaches.
The question of a static pair-gap in the $T\!=\!0$ channel can
be addressed meaningfully in the mean-field:
if the correlations are strong enough, it will show up
in a 'deformed' solution, otherwise not. This is of
course linked to the strength of the
interaction, which still needs to be determined.
Generalized BCS- and HFB-
calculations clearly indicate that
$T\!=\!0$ correlations are coherent and necessary 
to understand the experimental data\cite{Goo79,Sat97,Ter98}.

We are thus lead to the following conclusions:
i)the spin-orbit force in the nuclear potential quenches part
of the $L\!=\!0$ coupling and ii) the restriction to $L\!=\!0$ pairs only
takes into account a minor fraction of the real correlations. 
To proper investigate the full spectrum
of $T\!=\!0$, np-correlations, one
needs to take into account all J-values (implying all possible
L). The $T\!=\!0$ pairing as we define in the following, will have
contributions from all $J$'s. In an even-even nucleus e.g.
the spin $I\!=\!0^+$ state has a contribution from the standard
monopole-pairing with pairs of $J\!=\!0$ but of course also
the contribution from all odd $J-$values.
Note, however, that even the $\delta$-force strongly simplifies
the nucleon short-range interaction,
since it allows only the coupling of even $L$'s to take part.
The interaction related to space antisymmetric states
is not taken into account in any model involving a $\delta$-force
like interaction. 


\section*{Band termination at high angular momenta}

As discussed in the previous section, one may not see
obvious fingerprints of the $T\!=\!0$ pairing in the low spin regime,
since it is dominated by the breaking of pairs with low $J$ and 
with respect to that, there is little difference in $J\!=\!0$ and $J\!=\!1$.
In order to investigate the importance of the $T\!=\!0$ correlations
at high angular momenta, we have performed a case study for the
nucleus $^{48}$Cr\cite{Ter98}, that has been investigated quite extensively in
experiment and theory\cite{Le96,Ca96}. The valence particles of
$^{48}$Cr are placed in the middle of
the $f_{7/2}$ shell and the evolution of collective motion and
band termination is well described by shell-model calculations\cite{Cau95}.
The shell model faces difficulties to address states beyond the
fp-shell. Hence, mean-field calculations may shed some
light on the properties of states beyond the aligned $16\hbar^+$
state.

\begin{figure} [hbt] 
\vspace*{13.5cm}
\hspace*{2. cm}
\includegraphics{fig: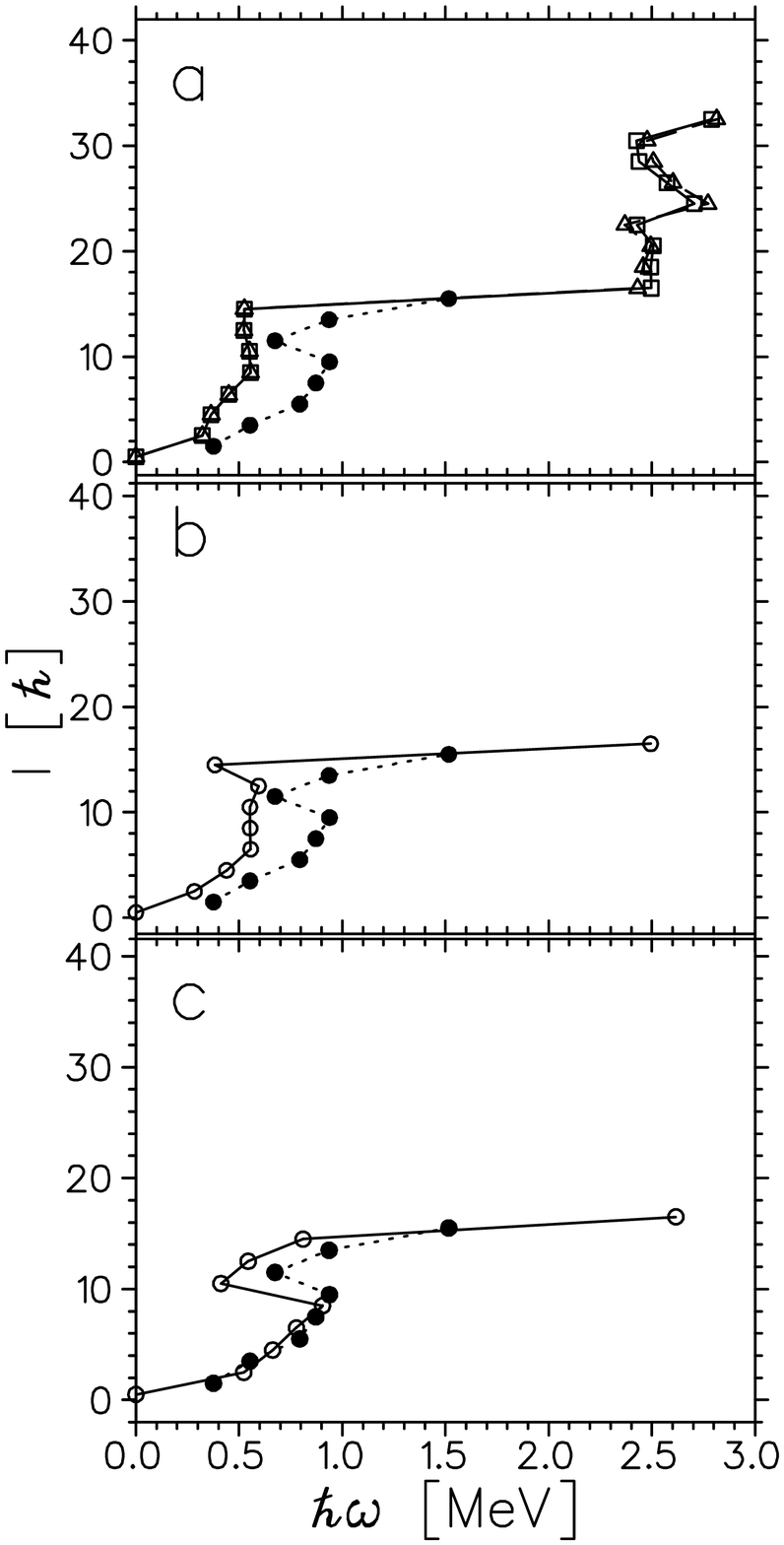}
\noindent
\hspace*{2.cm}
\vspace*{-.5cm}
\caption{Comparison of calculated and experimentally deduced
$I_x$ as a funciton of rotational frequency, $\hbar\omega$.
In a) we show two sets of 
calculations, with $G_{T0}\!=\!1.1G_{T1}$ and $1.3~G_{T1}$.
The low spin part, $I\le 16\hbar$ has only $T\!=\!1$ pairing,
wheras above that spin value, a transition to $T\!=\!0$ pairing occurs.
In b) the HFB-solution is dominated by $T\!=\!0$ pairing correlations, although
a small part of $T\!=\!1$ pairing is present simultanously.
c) The contribution of the $T\!=\!0$ pairing energy is added to 
the solution of a), according to $d\omega\!=\!\frac{dE_{pair}}{dI}$.
}
\label{fig2}
\end{figure}


We have performed cranked HFB-calculations in $R$-space based on the
Skyrme-force\cite{Bon87,Te95} where the two-body pp interaction
has been extended to incorporate both the $T\!=\!0$ and $T\!=\!1$ 
channel\cite{Ter98}.
The only symmetry restriction used in the calculations is
parity\cite{Ter98}. In the low spin regime, we found two HFB-solutions,
dominated by either the $T\!=\!0$ or $T\!=\!1$ correlations,
that were almost degenerate in energy, see a) and b) in Fig.\ref{fig2}.
Note, that for the case of the $T\!=\!0$ solution, both
pairing modes are present, showing the capability of our
approach to incorporate the two different pairing channels simultanously.
As shown in the previous section, the exact solution will
always mix both pairing modes
whereas the HFB approximation usually gives preference
to either one. As discussed in ref.~\cite{Sat97},
particle number projection results in mixed solutions
and one may expect that also iso-spin projection
will become necessary.  These steps will be considered 
in a further development of our model.

The results from the low-spin solutions
appear very interesting:
As seen in Fig.~\ref{fig2}, the gross
features of the alignment of the solution that is dominated by 
$T\!=\!0$ pairing (b) does not differ very much from the solution with
$T\!=\!1$ (a). This underlines the previous discussion - the low
spin regime will always be characterized by the pair-breaking
mechanism - be it $J\!=\!1$ or $J\!=\!0$ pairs. Nevertheless,
the slope of $I_x$ at low spins of the $T\!=\!0$ solution
somewhat better reproduces the experimental values.

The small value of the moments of inertia of atomic nuclei have
been taken as a fingerprint for the $T\!=\!1$ pairing
correlations, where the size of the reduction with respect to the value
of the rigid body indicates 
the strength of the correlations. Apparently, in the low spin regime,
we obtain a similar reduction for the $T\!=\!0$-pairing, where again the
size of the reduction is a measure of the correlations. Unfortunately,
one cannot disentagle what is the contribution due to
$T\!=\!0$ and $T\!=\!1$ pairing.
Since the two solution are exptected to
mix, we may artificially add the contribution of the 
pairing energy from the $T\!=\!0$-pairing  to the $T\!=\!1$-solution.
The result of such a mixing 
is shown in Fig.~\ref{fig2} c), resulting in a much improved
agreement with experiment. To really investigate the
effect of $T\!=\!0$ and $T\!=\!1$ pairing at low spins, one needs to go
beyond the mean-field description.

\begin{figure} [htb] 
\vspace*{13.0 cm}
\hspace*{3. cm}
\includegraphics{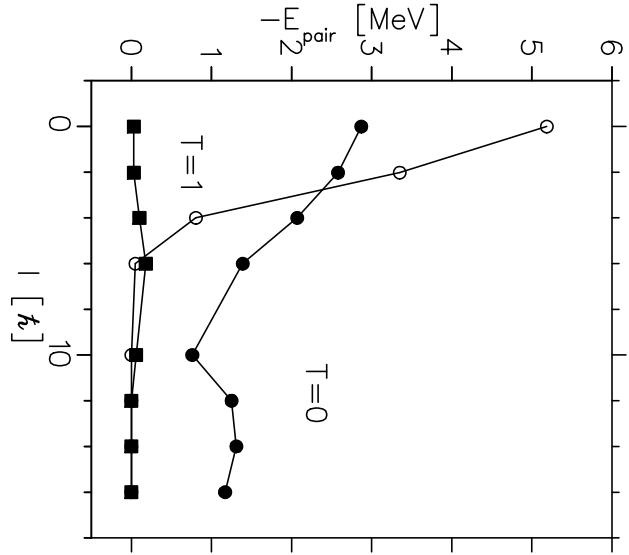}
\noindent
\hspace*{2.cm}
\vspace*{-4.cm}
\caption{The pairing energy of the $T\!=\!0$ (solid symbols)
and $T\!=\!1$ (open symbols)  Skyrme-HFB
solutions at low spins. Note the small value of the $T\!=\!1$ pairing
for the $T\!=\!0$ solution.}
\vspace*{10pt}
\label{fig3}
\end{figure}

In fig.\ref{fig3} the values of the pairing energy for
the two solutions are compared.
Indeed, the $T\!=\!0$ pairing is considerably more resistant
to rotation than the $T\!=\!1$.  The $T\!=\!0$ pairing energy
reveal a similar drop as the $T\!=\!1$, but it does not
approach zero and it increases again at the
$12\hbar$ aligned state, which has a shape close to spherical.
This dip in the pairing energy is related to the calculated
backbend. 
Also the intrinsic quadrupole moments
of the two solutions differ.  For transitional nuclei,
the difference can become quite large. The nucleus $^{44}$Ti e.g.
is calculated to have spherical shape in the presence of
$T\!=\!1$ pairing whereas it has a calculated quadrupole moment
of $\approx 50$eb for the $T\!=\!0$ solution.

The latter point is rather
important and we will discuss it
shortly. The pair-breaking mechanism is similar
for $T\!=\!0$ and $T\!=\!1$, and one may 
not expect strong influence on the
band-crossing frequencies. Nevertheless, since
the $T\!=\!0$ pairing field 
affects the shape of the nucleus, and band-crossing frequencies
quite sensitively depend on deformation\cite{Wys89}, the two effects are linked
together. 
For the case of $^{72}$Kr e.g., 
a large delay of the band crossing has been observed\cite{Ang97}.
A possible mechanism
emerging from our result would be that the nucleus stays at a more
deformed shape, thus delaying the crossing frequency.
In addition, 
since in real nuclei more states
than a single j-shell contributes to
the pairing energy, one expects more
binding from all different $J\!=\!1$ states, 
resulting in additional ground-state correlations.
\begin{figure} [htb] 
\vspace*{13. cm}
\hspace*{3. cm}
\includegraphics{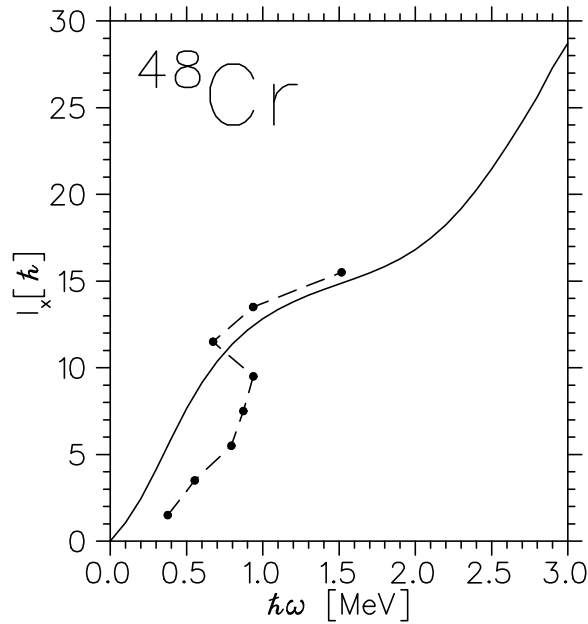}
\noindent
\hspace*{2.cm}
\vspace*{-4.cm}
\caption{Cranked Woods-Saxon calculations for $^{48}$Cr.
The calculations are done at spherical shape and allow for both
$T\!=\!0$ and $T\!=\!1$ pairing. The $T\!=\!0$ pairing field allows for a 
smooth
occupation of the aligned deformation driving intruder orbitals}
\vspace*{10pt}
\label{fig4}
\end{figure}

For the high spin regime, we find only a solution with finite
$T\!=\!0$ gap. This result clearly demonstrates
the onset of $T\!=\!0$ pairing at high angular momenta,
since the static $T\!=\!1$ correlations are quenched
at these spin-values.
The $T\!=\!0$ pairing energy rises after the terminating
$I=16\hbar$-state and maximizes at $I=24\hbar$, where it has
approximatively the same value as the nn (pp) $T\!=\!1$ 
pairing at $I=0\hbar$ (2.2~MeV). The size of the $T\!=\!0$ 
correlation implies the presence
of a static $T\!=\!0$ pairing field.
The increase in angular momenta 
beyond the terminating $I\!=\!16\hbar$ state
is due to excitation from $d_{3/2}$ and $f_{7/2}$ into $g_{9/2}$
and $f_{5/2}$ orbits. Note that these particle-hole
excitations are caused by the short range $T\!=\!0$
pairing force. Without the $T\!=\!0$ correlations, no cranked HFB- 
or HF-solutions are
found in the region between $I\!=\!16$ and $I\!=\!32\hbar$
\cite{Ter98}.

The effect of the $T\!=\!0$ pairing
is to allow for a smooth occupation of the aligned
\hbox{$[(d_{3/2})^{-2}_{3^+}(f_{7/2})^{-2}_{+1}-
(f_{5/2})^2_{+5}(g_{9/2})^2_{9^+}]_{18^+}$}
configuration (the $(f_{7/2})^{-2}$ holes are with
respect to the $(f_{7/2})^8_{16^+}$ configuration).
The occupation probability 
of this configurations is increasing
smoothly, resulting in an almost constant increase in angular momenta 
and deformation
as a function of frequency.
Since the configuration 
is exactly half filled at $I\!=\!24\hbar$, 
we obtain a maximum
for the pairing correlations at that spin value\cite{Ter98}. 

Note that this 'vertical' excitation is completely different
to the common ph-excitations, that are associated with
the shape coexisting band structures. 
One may e.g. promote
2 particles from the $d_{3/2}$ into the $g_{9/2}$. This 2p-2h
excitation couples to angular momentum $I\!=\!0\hbar$ and may result in a 
new rotational band. In this case, the $g_{9/2}$ is fully occupied
and the $d_{3/2}$ is empty and the angular momentum gain arises
from the alignment of the valence particles. This kind of
'horizontal' ph-excitation form the well-know intruder bands or
highly- and super-deformed band structures.
In the case of the $T\!=\!0$ pairing, one instead smoothly
occupies the aligned configuration, resulting in a smooth
deformation change.
This presents a new kind of collective
excitations, where the $T\!=\!0$ pairing field induce
an increase in deformation and angular momenta.

\begin{figure} [htb] 
\vspace*{13. cm}
\hspace*{3. cm}
\includegraphics{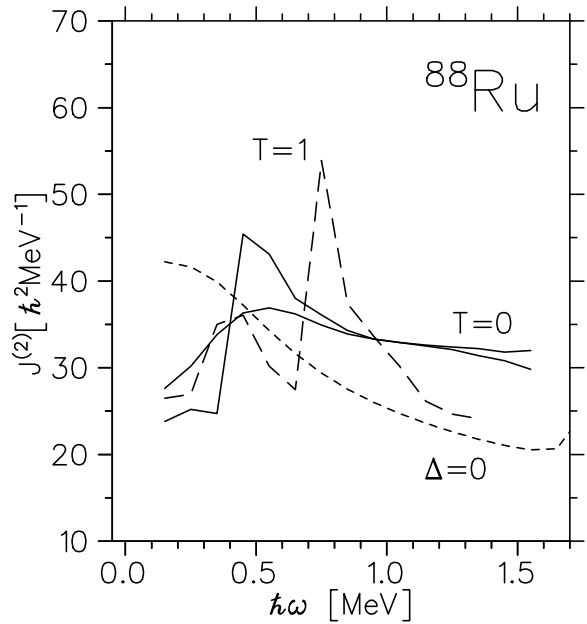}
\noindent
\hspace*{2.cm}
\vspace*{-4.cm}
\caption{The $J^2$ moment of inertia
resulting  from the minimum of the TRS-calculations with different
parameters of the 
pairing correlations. 
For further information, see text.}
\vspace*{10pt}
\label{fig5}
\end{figure}

The same effect is seen in cranked Woods-Saxon calculations\cite{Sat99}.
Here $T\!=\!0$ and $T\!=\!1$ pairing are present simultanously and 
contribute to the total energy. Approximate particle number projection
is performed by means of the Lipkin Nogami procedure. More details can be
found in \cite{Sat97,Sat99}. An important difference of the
cranked Woods-Saxon calculations is the lack of feed back 
from the ocupation of intruder orbitals to the
single
particle potential.
Although the same configurations are occupied as in the cranked Skyrme-HFB
calculations, the shape stays spherical, since the single particle
potential does not feel the intrinsic deformation of the ph-excitations.
Remarkably, the presence of $T\!=\!0$ pairing allows for a smooth increase
of the angular momenta at spherical shape. Since we are cranking
around a symmetry axis, 
this is in contrast to the
step wise increase which is obtained for normal cranking with or
without $T\!=\!1$, where each step 
in angular momentum is associated to the 
$<j_z>$ expectation value of each particular pair.
The presence of $T\!=\!0$ pairing allows instead a smooth 'filling' of the
aligned orbital. Again, we encounter a new mechanism to create
angular momentum and deformation. 

The feature discussed in this context is a general property of nuclei in the
close 
vicinity of $N\!=\!Z$. Cranked HFB calculations for $^{44}$Ti e.g. show the same
mechanism. After the terminating state at $I\!=\!12\hbar$,
$T\!=\!0$ correlations start to show up, resulting in a further increase
of the angular momenta. 
For the case of $^{44}$Ti, the $T\!=\!0$-solution is considerably
more deformed than the spherical $T\!=\!1$-solution. Again,
we expect only one solution to be physical, but in the
presence of $T\!=\!0$ correlations, there is a clear
tendency to more deformed shapes, that shows up
especially in transitional nuclei.
In general, there will always be the
mechanism to go beyond the terminating state by means of
$T\!=\!0$ induced ph excitations. These 'vertical' excitations will
of course have to compete with the 'horizontal' excitations, where
by deforming the nucleus a particular set of favoured ph-excitations
can lead to a more deformed shape and result in a new collective rotational
band.

An interesting situation may arise at super-deformed shape in
$N\!=\!Z$ nuclei. Cranked Woods-Saxon calculations predict
very deep minima at super- and hyperdeformed shapes in
$^{88}$Ru and $^{92}$Pd. The
very stable shapes of these doubly magic super- and hyper-deformed
nuclei have similar configuration as the harmonic oscillator
at 2:1 shape for $N\!=\!Z\!=\!40$. The experimental investigation of
this shell structure is still in progress, but we may expect with the
event of radio-active beam facilities to probe such exotic nuclei.
Since the valence shell is particular large for super-deformed
nuclei, the fingerprints of $T\!=\!0$ type pairing may become enhanced,
especially at high angular momenta. 

Results of our
total routhian surface calculations are shown in Fig.\ref{fig5}.
We compare four different sets of calculations:
The standard solution where we allow for
$T\!=\!1$ pairing only (dashed line), shows a sharp crossing related to the
alignment of $h_{11/2}$ protons and neutrons. After that bandcrossing,
the moment of inertia drops and approaches the curve, where no
pairing correlations are present ($\Delta\!=\!0$, short-dashed line).
The curves shown with solid lines correspond to a calculation,
where we allow both $T\!=\!0$ and $T\!=\!1$ pairing to be present.
The strength of the $T\!=\!0$ pairing field is scaled with respect to
$T\!=\!1$, see the discussion in \cite{Sat97}. The curve that 
shows only a smooth hump in the frequency range of $\hbar\omega\approx0.5$~MeV
has a strenght of $G_{T\!=\!0}\!=\!1.3G_{T\!=\!1}$, implying that the
$T\!=\!0$ pairing field is present already at $\hbar\omega\!=\!0.0$~MeV.
No sharp band crossing is observed for that case.
When we choose a strength that is undercritical,
$G_{T\!=\!0}\!=\!1.1G_{T\!=\!1}$, implying that the pairing is of
$T\!=\!1$ type at $\hbar\omega\!=\!0.0$~MeV, we experience a
pairing phase transition from $T\!=\!1$ to $T\!=\!0$, which
is related to the sharp crossing at $\hbar\omega\!=\!0.5$~MeV.
This kind of transition has been discussed previously
in Ref.~\cite{Mul81}.
and reflects the mean-field approximation.

Note that the moment of inertia for a rotational band, where $T\!=\!0$ pairing
correlations are present, exceeds by a far amount the value
that is obtained for 'rigid' rotation (i.e. no pairing
correlations). The nucleus appears thus more 'rigid' in
the presence of $T\!=\!0$ correlations.
This is exactly the same phenomenon as discussed above concerning
the band-termination. Since the angular momentum in a nuclues
is built up from the contribution of individual nucleons,
the angular momentum space is limited for each specific configuration.
This is in contrast to a rigid body, where there is in principal
no limitation, beyond fission. In the presence of $T\!=\!0$ pairing,
the configuration space is not limited anymore, since now the short range
correlations will always scatter pairs into orbitals with larger
angular momenta, and hence allow for an increase in angular momentum.
Hence, rotation in that case really resembles the rotation of a rigid
body, and the drop in the moment of inertia, which is obtained in
all nuclei at a certain frequency, may not be observed.

\section*{Conclusions.}

We have presented different model studies that explore
the iso-spin degree of freedom in the particle-particle channel.
The $T\!=\!0$ pairing correlations are expeted
to influence the structure of $N\!=\!Z$ nuclei.
In order to discuss these correlations, one certainly cannot
restrict to the narrow definition of $L\!=\!0$ pairing.
As shown for the case of a single j-shell, all $L$-values
contribute to the correlations.
In the context of the mean-field, it is clear that a
$T\!=\!0$ pairing gap is obtained, i.e. the correlations
are coherent. The $T\!=\!0$ pairing is resistant to rotation
and modifies the rotational spectrum at high angular velocities.
A new kind of collective excitation mode appears, allowing for
smooth and continous occupation of high-j orbits.


\section*{Acknowledgements.}

This work is done in collaboration with
P.-H. Heenen, U.L.B Brussels, W. Satu{\l}a, Univ. Warsaw and
KTH, 
J. Sheikh, KTH and
Tata Institute of Fundamental Research, 
and J. Terasaki, KTH.
Discussions with J. Blomquist are acknowledged. 
We are thankful to the support given
by the G\"oran Gustaffsson foundation,
the Swedish Institute, the Swedish Natural Research Council (NFR)
and the Axel och Margaret Ax:son Johnson Stiftelse.

\end{document}